\documentclass{iaus}
\usepackage{graphicx}

\title[3D-MHD simulations of evolving of magnetic fields in FR~II sources]
{3D-MHD simulations of the evolution of magnetic fields in FR~II radio sources}

\author[Huarte-Espinosa, Krause \& Alexander]
{Mart\'{\i}n Huarte-Espinosa$^{1,3}$, Martin Krause$^{4,5}$
 \and Paul Alexander$^{2,3}$}

\affiliation{$^1$Department of Physics and Astronomy, University of Rochester
600 Wilson Boulevard, Rochester, NY, 14627-0171;~ 
$^2$Astrophysics Group, Cavendish Laboratory, 19~J.~J. Thomson~Ave.,
Cambridge CB3 0HE, UK \\
emails: {\tt martinhe@pas.rochester.edu, pa@mrao.cam.ac.uk}\\
$^3$Kavli Institute for Cosmology Cambridge, Madingley Road, Cambridge CB3 0HA,~UK;
~$^4$Max-Planck-Institut f\"ur Extraterrestrische Physik,
Giessenbachstrasse, 85748 Garching,~Germany.
email: {\tt krause@mpe.mpg.de} \\
$^5$Universit\"atssternwarte M\"unchen, Scheinerstr.~1, 81679 M\"unchen, Germany
}
\pubyear{2011}
\volume{275}
\pagerange{119--126}
\jname{Jets at all Scales}
\editors{A.C. Editor, B.D. Editor \& C.E. Editor, eds.}
\begin{document}

\maketitle

\begin{abstract}
3D-MHD numerical simulations of bipolar, hypersonic, weakly magnetized
jets and synthetic synchrotron observations are presented to study
the structure and evolution of magnetic fields in FR~II radio
sources. The magnetic field setup in the jet is initially random.
The power of the jets as well as the observational viewing angle are
investigated.  We find that synthetic polarization maps agree with observations
and show that magnetic fields inside the sources are shaped by the
jets' backflow.  Polarimetry statistics correlates with time, the
viewing angle and the jet-to-ambient density contrast. The magnetic
structure inside thin elongated sources is more uniform than for
ones with fatter cocoons. Jets increase the magnetic energy in
cocoons, in proportion to the jet velocity.  Both, filaments in
synthetic emission maps and 3D magnetic power spectra suggest that
turbulence develops in evolved sources.
\end{abstract}

\firstsection 
\section{Introduction}

Fanaroff-Riley class~II radio sources (FRIIs, \cite[Fanaroff \&
Riley 1974]{fr}) are extragalactic, synchrotron in nature and show
linear polarization fractions within 10--50\% (\cite[Bridle \&
Perley 1984]{bridle}). Stokes parameters are used to infer the
magnetic structure in these sources. Observed magnetic polarization
vectors are generally parallel to the jets and to the lobe boundaries, and follow
both flux intensity gradients perpendicularly and lines between
multiple lobe hot~spots (\cite[Bridle \& Perley 1984]{bridle}).
The linear polarization fraction in FRIIs is typically higher at jet edges
than in their beams, and also at source edges than in the cocoons
(\cite[Saikia \& Salter 1988]{saikia}). The magnetic structure in
FRIIs, as well as the way it evolves and relates to AGN jet properties,
is not clear.

\section{Model and methodology}
The equations of ideal MHD are solved in 3D using the code Flash~3.1
(\cite[Fryxell B. et al. 2000]{flash}), inside a cubic Cartesian domain
with \hbox{200$^3$ fixed cells}.
The intra-cluster medium is implemented as a monoatomic ideal gas
($\gamma=5/3$), a stratified King density profile (\cite[King
1972]{king}), magnetohydrostatic equilibrium with a central
gravitational field and magnetic fields with a Kolmogorov turbulent
structure, with a thermal-to-magnetic pressure ratio~$\gtrsim\,$10.
Source terms are implemented in the equations
to inject mass and \hbox{$x$-momentum} in a central cylinder
which takes weak and random magnetic fields from the innermost
ambient medium. We investigate jet velocities with Mach=\{40, 80,
130\} as well as $\eta=\rho_{\mbox{jet}}/\rho_{\mbox{amb}}=$ \{0.01,
0.001\}.

\section{Synthetic synchrotron emission}
Synchrotron emission and Stokes parameters are calculated and
integrated through the inflated model sources, along the line of
sight.  The density distribution of ultra-relativistic electrons
is the product of the cocoon pressure and an incompressible tracer
field injected with the jets.  Synthetic polarization maps are
produced for five model sources at different source expansion times,
$t_{\mbox{jet}}$, and for viewing angles $\theta_v=\,$\{30$^{\circ}$,
60$^{\circ}$, 90$^{\circ}$\} (e.g. Figure~1).

\begin{figure}[ht]
\begin{center}
 \includegraphics[width=0.359\columnwidth,bb=-.5in 2.5in 4.87in 9.25in,clip=]{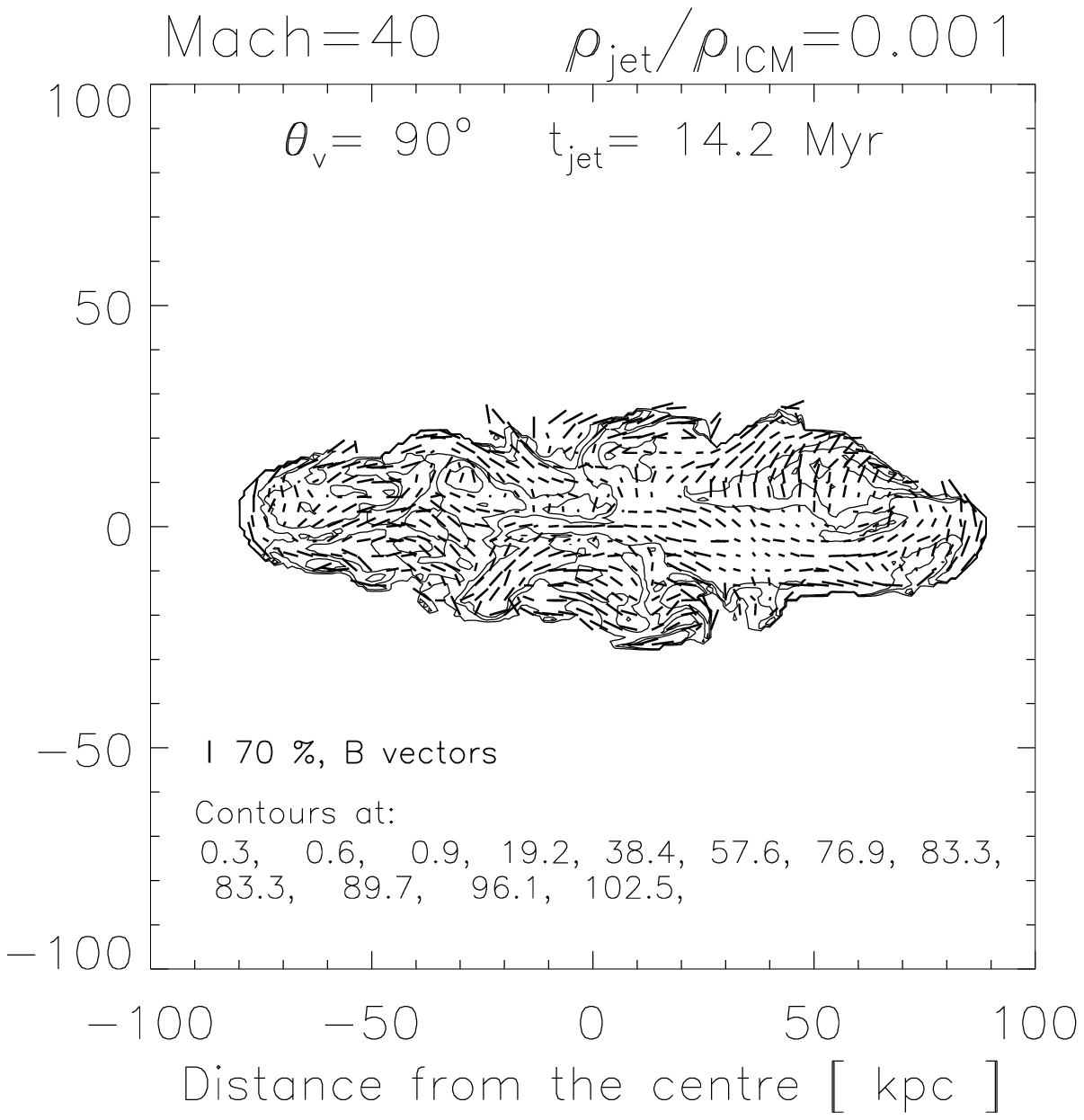} ~~~~
 \includegraphics[width=0.313\columnwidth,bb=.15in 2.49in 4.87in 9.25in,clip=]{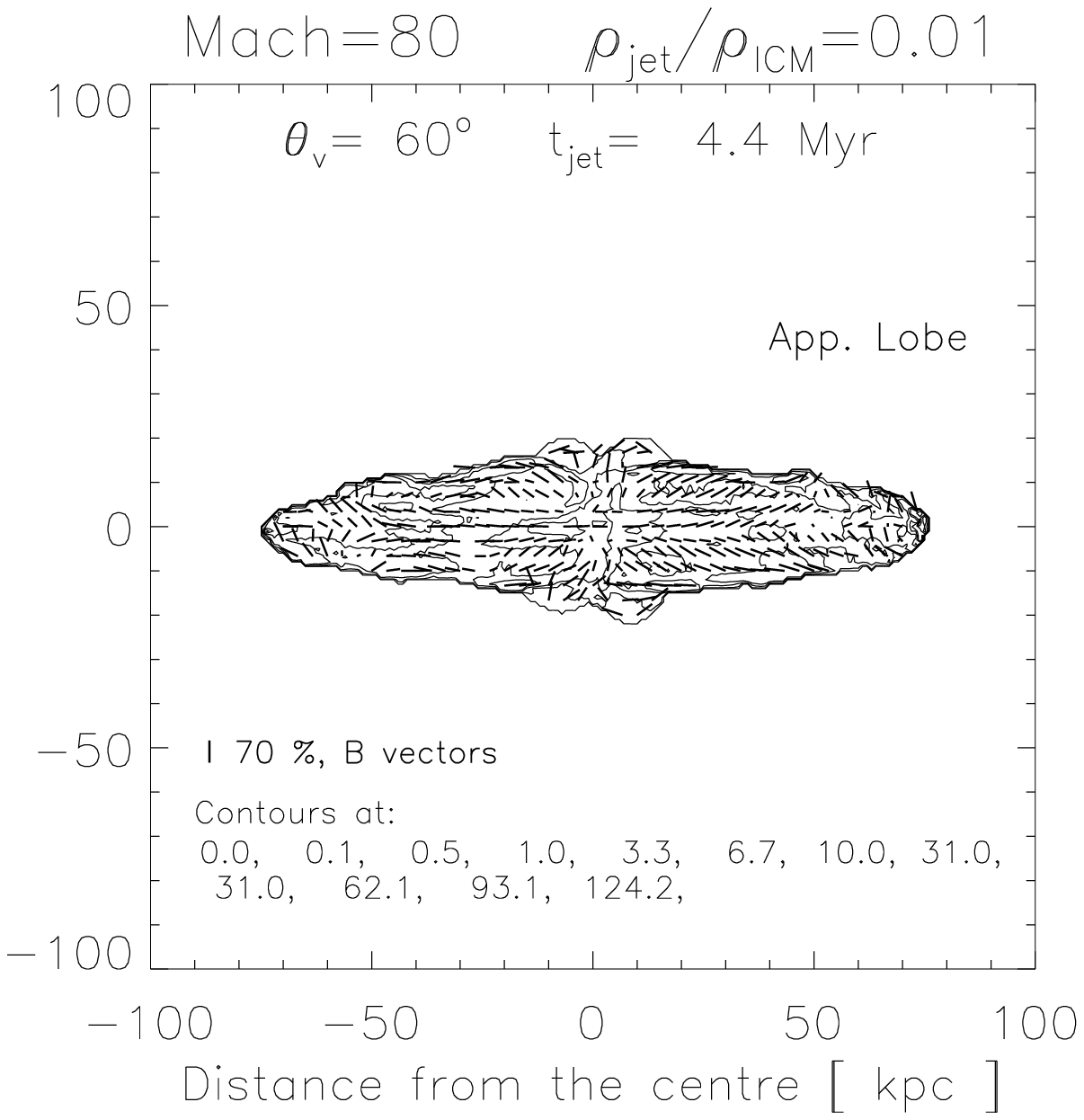}
  \vspace*{ -1.75 cm}
\caption{Synthetic polarization maps. Vectors follow the magnetic
position angle and their length is proportional to the degree of
linear polarization. Vectors are superimposed on contours
of synchrotron emission (at 8\,GHz) normalized to the mean emissivity.
The initially random magnetic fields have been ordered by MHD processes.
The left lobe in the right panel source is receding, 
yet beaming and light-travel effects are assumed to be negligible.
}
\end{center} 
\label{fig1} 
\end{figure}

\section{Conclusions}

Jets injected with initial random magnetic fields develop ordered
fields by MHD processes within the radio source.
Filaments suggest that turbulence develops in evolved sources.
Polarimetry statistics correlates with time, $\theta_v$ and $\eta$,
but not so with v$_{\mbox{jet}}$. Lighter jets show linear polarization
degrees~$\sim \,$39\%
at the end of the simulations, independently of $\theta_v$. This
agrees with observations better than for the heavier sources which
show better, and more realistic, field alignment
with the jets than lighter sources. Some initial order 
in the magnetic fields may be
required to meet all the constraints at once. See 
Huarte-Espinosa, Krause \& Alexander~2011a (in prep.) for details.\\

{\scriptsize The software used in these investigations was in part
developed by the DOE-supported ASC / Alliance Center for Astrophysical
Thermonuclear Flashes at the University of Chicago. MHE acknowledges
funding from CONACyT M\'exico 196898/217314, and Dongwook~Lee for the
3D-USM-MHD solver of Flash~3.1.}

\begin{discussion}

\end{discussion}

\end{document}